\documentclass[10pt,letterpaper,twocolumn]{revtex4-1} %% two column, final layout

\usepackage{amsmath}
\usepackage{amssymb}
\usepackage{graphicx}
\usepackage{epstopdf}
\usepackage{bbm}
\usepackage{mathrsfs}
\usepackage[dvipdfm,colorlinks=true,pdfstartview=FitV,linkcolor=blue,citecolor=blue,urlcolor=blue]{hyperref}

\newcommand{\diff}{\mathrm{d}}
\newcommand{\blue}[1]{\textcolor{blue}{#1}}

\begin{document}

\title{Quantum entanglement in plasmonic waveguides \\
       with near-zero mode indices}
\author{Xing Ri Jin, Lei Sun, Xiaodong Yang, and Jie Gao$^{\blue{*}}$}
\address {Department of Mechanical and Aerospace Engineering, \\
Missouri University of Science and Technology, Rolla, Missouri 65409, USA \\
Corresponding author: $^{\blue{*}}$gaojie@mst.edu}
%%%%%%%%%%%%%%%%%%%%
%%%% Abstract %%%%%%
%%%%%%%%%%%%%%%%%%%%
\begin{abstract}
We investigate the quantum entanglement between two quantum dots in
a plasmonic waveguide with near-zero mode index, considering the dependence
of concurrence on interdot distance, quantum dot-waveguide frequency detuning
and coupling strength ratio.
High concurrence is achieved for a wide range of interdot distance due to
the near-zero mode index, which largely relaxes the strict requirement of interdot distance
in conventional dielectric waveguides or metal nanowires.
The proposed quantum dot-waveguide system with near-zero phase variation
along the waveguide near the mode cutoff frequency shows very promising
potential in quantum optics and quantum information processing.
\end{abstract}

\maketitle

%%%---------------%%%%
% Introduction
%%%---------------%%%%

Highly entangled quantum states play important roles in quantum information
science, such as schemes for quantum cryptography, quantum teleportation,
and quantum computation~\cite{Ekert1991, Bennett1993, Ritter2013, Knill2001}.
Among different physical realizations, scalable solid-state quantum entanglement
is the most promising one, and the recent successes of quantum entanglement have
been obtained with quantum dots (QDs) or diamond nitrogen vacancy
centers~\cite{Gossard2005, Gershoni2006, Lukin2010, Hanson2013}
in the visible frequency range.
For long-distance entanglement, the correlation between two spatially separated qubits
is usually mediated by photons.
However, instead of photon, surface plasmon~\cite{Raether1988} has been attracted
much attention since it reveals strong analogy to light propagation in conventional
dielectric optical components.
Plasmonic waveguides and resonators can be used to confine light to subwavelength
dimensions below the diffraction limit for achieving photonic circuit
miniaturization~\cite{Ebbesen2006} and furthermore strongly interact with quantum emitters
for the applications of detectors, transistors and quantum information
processing~\cite{Chang2006, Chang2007,ChenGYOL1337,Jin2013}.
For one-dimensional plasmonic waveguide, the scattering properties of surface
plasmon interacting with QDs have been studied
widely~\cite{Akimov2007,Fedutik2007,Weinano2009,Chenoe2010,Kimapl2010,ChenGYOL4023}.
Recently, Chen \textit{et al.}~\cite{ChenGYPRB045301}
and Gonzalez-Tudela \textit{et al.}~\cite{Gonzalez2011}
have reported quantum entanglement generation between two separated QDs mediated
by a plasmonic waveguide.
Highly entangled state between QDs can be achieved only when the interdot distance
is controlled with specific values, due to the sinusoidal phase variation of the
propagating surface plasmon mode in the waveguide.

%%---%%
In this work, we examine plasmonic waveguides with near-zero mode indices,
and investigate the quantum entanglement between
two QDs simultaneously interacting with the waveguide mode.
High concurrence of the entangled state can be obtained for wide ranges of
interdot distance $d$ and QD-waveguide coupling strength ratio $g_{2}/g_{1}$,
showing great advantage of relaxing the strict
requirements of QDs positions in comparison to previous schemes.
The physical mechanism of the interdot distance flexibility is the vanishing phase variation
between two arbitrary positions along the plasmonic waveguide with near-zero mode index.
With the pioneering experimental verification of $n = 0$ structures for
visible light by Vesseur\textit{et al.}~\cite{VesseurPRL110},
the proposed QD-waveguide platform is convinced to be a promising
experimental platform for realizing highly entangled quantum states.

%- FIGURE 1 -%%
\setlength{\intextsep}{8pt plus 2pt minus 2pt}
\begin{figure}[htbp]
    \centering
    \includegraphics[width=8.0cm]{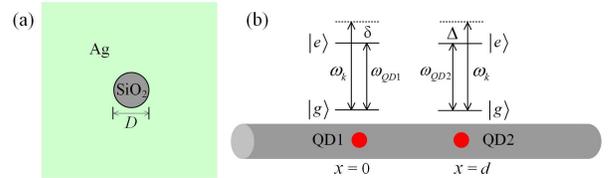}
    \vspace{-10pt}
    \caption{(Color online)
        (a) Schematic of the cross section of a $\mathrm{SiO}_{2}$
        waveguide with a thick silver cladding.
        $D$ is the diameter of $\mathrm{SiO}_{2}$ core.
        $3\,\mu\mathrm{m}$-long waveguide is used in the model.
        (b) Two two-level QDs separated by distance $d$ interacting
        with the waveguide mode.
        $\delta$ ($\Delta$) is the frequency detuning between $\mathrm{QD}_{1}$
        ($\mathrm{QD}_{2}$) transition and the incident waveguide mode.}
\end{figure}
%- END -%%

%%---%%
%We consider a metal-coated dielectric waveguide interacting with two
%embedded QDs in the model.
%%
Fig.~\blue{1(a)} shows a schematic of one $\mathrm{SiO}_{2}/\mathrm{Ag}$ waveguide engineered
to exhibit near-zero mode index in the visible frequency range.
The diameter of the $\mathrm{SiO}_{2}$ core is $D$, which is fully surrounded
by a thick silver cladding.
The permittivity of silver is described by the Drude model, with the plasmon
frequency $\omega_p$ of $1.37\times10^{16}\,\mathrm{rad}/\mathrm{s}$ and
the collision frequency $\gamma$ of $8.5\times10^{13}\,\mathrm{rad}/\mathrm{s}$
at room temperature.
The refractive index of $\mathrm{SiO}_{2}$ is $1.46$.
A pair of QDs, each of which has one excited state $\left|e\rangle\right.$ and
one ground state $\left|g\rangle\right.$, are embedded in the $\mathrm{SiO}_{2}/\mathrm{Ag}$
waveguide as illustrated in Fig.~\blue{1(b)}.
$\delta$($\Delta)=\omega_{k}-\omega_{1(2)}$ is the frequency detuning between
the incident waveguide mode and the QD exciton transition.
$d$ is the interdot distance between the two separated QDs.

%%---%%
The mode indices $n$ of $\mathrm{SiO}_{2}/\mathrm{Ag}$ plasmonic waveguides
are plotted in Fig.~\blue{2(a)} where the loss of $\mathrm{Ag}$ is neglected.
Near-zero mode index can be reached around the cut-off frequency of the waveguide mode,
which can be designed by varying the waveguide diameter.
The working wavelength with near-zero mode index can be controlled from
visible $685\,\mathrm{nm}$ to near-infrared $920\,\mathrm{nm}$ when the
waveguide diameter $D$ changes from $100\,\mathrm{nm}$ to $150\,\mathrm{nm}$.
Regarding the experimental condition at cryogenic temperature for
the interaction between visible QDs and waveguide mode, silver cladding
layer with $0.1\gamma$ damping rate and waveguide of $D = 110\,\mathrm{nm}$
are considered in the following analysis.
Fig.~\blue{2(b)} shows the corresponding mode index $n$ and group velocity $v_{g}$,
where the mode index gradually approaches to a vanishing small number and group
velocity slows down to $c/42$ at the wavelength of $728.6\,\mathrm{nm}$.
Group velocity $v_{g}$ is calculated according to $v_{g}= c/(n+\omega (dn/d\omega))$,
where $c$ is the speed of light in vacuum.
Figs.~\blue{2(c)} and \blue{2(d)} show electric field distributions at the wavelengths
of $600\,\mathrm{nm}$ and $725\,\mathrm{nm}$ for the $\mathrm{SiO}_{2}/\mathrm{Ag}$
waveguide calculated in Fig.~\blue{2(b)}.
The corresponding mode indices at the wavelengths of $600\,\mathrm{nm}$
and $725\,\mathrm{nm}$ are $0.962$ and $0.164$, respectively.
For the waveguide mode with near-zero index, light can propagate along
the waveguide with a spatially uniform phase, near infinity phase velocity
and slow group velocity.

%- FIGURE 2 -%%
\setlength{\intextsep}{4pt plus 2pt minus 2pt}
\begin{figure}[htbp]
    \centering
    \includegraphics[width=5.25cm]{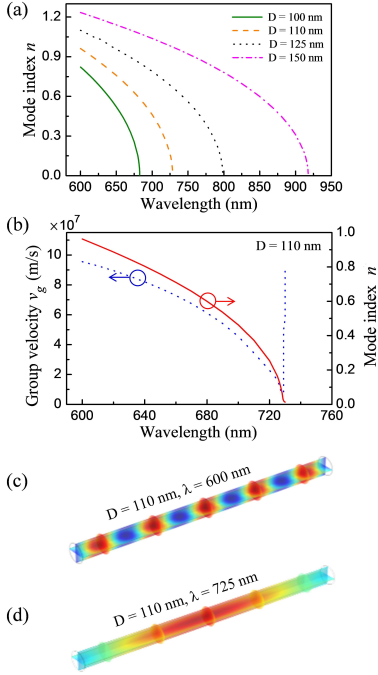}
    \vspace{-10pt}
    \caption{(Color online)
        (a) Dependence of waveguide mode indices $n$
        on the wavelengths for different diameters $D$ at
        $100\,\mathrm{nm}$ (green solid line),
        $110\,\mathrm{nm}$ (orange dashed line),
        $125\,\mathrm{nm}$ (navy dotted line), and
        $150\,\mathrm{nm}$ (magenta dash-dotted line).
        (b) Dependence of group velocity $v_{g}$ and mode index $n$ on
        the wavelengths for the waveguide with $D = 110\,\mathrm{nm}$
        when $0.1\gamma$ $\mathrm{Ag}$ loss is considered.
        (c--d) The electric field distributions at the wavelengths
        of $600\,\mathrm{nm}$ and $725\,\mathrm{nm}$ for the waveguide
        discussed in (b).}
\end{figure}
%- END -%%

%%---%%
Here, we consider the incident waveguide mode with energy $E_{k}=v_{g}k$
interacting simultaneously with two QDs, where $k$ is the wave vector of
the incident mode.
Thus the real-space Hamiltonian can be written as (assuming $\hbar = 1$)
\begin{equation}
\setlength{\abovedisplayskip}{1pt}
\setlength{\belowdisplayskip}{1pt}
\label{eq:eq-one}
\begin{split}
H = & \int \diff x
       \left\{
        - iv_{g}c_{R}^{\dag}(x)\tfrac{\partial}{\partial x}c_{R}(x)
        + iv_{g}c_{L}^{\dag}(x)\tfrac{\partial}{\partial x}c_{L}(x) \right. \\
    & \left. + \sum^{2}_{j=1}g_{j}\delta[x-(j-1)d]
              [c_{R}^{\dag}(x)\sigma_{-}^{j}+c_{R}(x)\sigma_{+}^{j} \right. \\
    & \left. + c_{L}^{\dag}(x)\sigma_{-}^{j}+c_{L}(x)\sigma_{+}^{j}] \right\}
              + \sum^{2}_{j=1}(\omega_{j}-i\frac{\Gamma}{2})\sigma_{e_{j},e_{j}},
\end{split}
\end{equation}
where $c_{R}^{\dag}(x)$ [$c_{L}^{\dag}(x)$] is the bosonic
operator creating a right-going (left-going) surface plasmon at
the position $x$, and $g_{j}$ $(j=1,2)$ is the coupling strength between individual QDs
and the waveguide mode. Here, the dipole moments of the two QDs have the same orientation
and both dipole moments are parallel to the polarization direction of the waveguide mode.
$\sigma_{e_{j},e_{j}}=|e\rangle_{j}\langle e|$ represents the
diagonal element of the $j$th QD operator and
$\sigma_{+}^{j}=|e\rangle_{j}\langle g|$
$(\sigma_{-}^{j}=|g\rangle_{j}\langle e|)$ represents the rasing
(lowering) operator. %%
$\omega_{j}$ is the transition frequency of the $j$th QD and
$\Gamma$ is the total dissipation including the exciton decay to free space, ohmic loss and other dissipative channels.
The eigenstate of the system can be written as
\begin{equation}
\setlength{\abovedisplayskip}{2pt}
\setlength{\belowdisplayskip}{4pt}
\label{eq:eq-two}
\begin{split}
    |E_{k}\rangle =
        & \int \diff x [\phi_{k,R}^{+}(x)c_{R}^{\dag}(x)\\
        & +\phi_{k,L}^{+}(x)c_{L}^{\dag}(x)]
         |g_{1},g_{2}\rangle|0\rangle_{\rm{sp}} \\
        & + \xi_{k_{1}}|e_{1},g_{2}\rangle|0\rangle_{\rm{sp}}
         + \xi_{k_{2}}|g_{1},e_{2}\rangle|0\rangle_{\rm{sp}},
\end{split}
\end{equation}
where $\xi_{k_{1}}$ ($\xi_{k_{2}}$) is the probability amplitude that $\mathrm{QD}_{1}$
($\mathrm{QD}_{2}$) absorbs the waveguide mode and jumps to its excited state.
Supposing a surface plasmon incident from the left, the scattering amplitudes can be written as
$\phi_{k,R}^{+}(x)\equiv e^{ikx}[\theta(-x) + a\theta(x)\theta(d-x) + t\theta(x-d)]$ and
$\phi_{k,L}^{+}(x)\equiv e^{-ikx}[r\theta(-x)+b\theta(x)\theta(d-x)]$. $\theta(x)$
is the unit step function, which equals unity when $x\geqslant0$ and zero when $x<0$.
$a$ and $b$ are the probability amplitudes of the field between
the two QDs at $x = 0$ and $x = d$.
$t$ and $r$ are the transmission and reflection amplitudes, respectively.
By solving the eigenvalue equation $H|E_{k}\rangle=E_{k}|E_{k}\rangle$,
one can obtain the following relations for the coefficients:
\begin{equation}
\setlength{\abovedisplayskip}{3pt}
\setlength{\belowdisplayskip}{3pt}
\label{eq:eq-three}
\begin{aligned}
    & g_{1}(1+a+r+b) = (\delta+\frac{i\Gamma}{2})\xi_{k_{1}}, \\
    & g_{2}(2ae^{ikd}+2be^{-ikd})=(\Delta+\frac{i\Gamma}{2})\xi_{k_{2}}, \\
    & a=1+\frac{g_{1}\xi_{k_{1}}}{iv_{g}},\quad t=1+\frac{1}{iv_{g}}(g_{1}\xi_{k_{1}}+g_{2}\xi_{k_{2}}e^{-ikd}), \\
    & b=\frac{g_{2}\xi_{k_{2}}}{iv_{g}}e^{ikd},\quad r=\frac{1}{iv_{g}}(g_{1}\xi_{k_{1}}+g_{2}\xi_{k_{2}}e^{ikd}),
\end{aligned}
\end{equation}
Through solving the Eq.~\eqref{eq:eq-three},
$\xi_{k_{1}}$ and $\xi_{k_{2}}$ can be obtained as follows:
\begin{equation}
\setlength{\abovedisplayskip}{3pt}
\setlength{\belowdisplayskip}{3pt}
\label{eq:eq-four}
\begin{aligned}
  \xi_{k_{1}} &= \frac{-i4g_{1}[-4(-1+e^{2ikd})J_{2}+(\Gamma-2i\delta)]}{\eta},\\
  \xi_{k_{2}} &= \frac{-i4g_{2}e^{ikd}(\Gamma-2i\Delta)}{\eta}, \\
         \eta &= -16(-1+e^{2ikd})J_{1}^{2}J_{2}^{2}+4[J_{1}(\Gamma-2i\delta)\\
              &+ J_{2}(\Gamma-2i\Delta)]+(\Gamma-2i\delta)(\Gamma-2i\Delta),
\end{aligned}
\end{equation}
where $J_{1}=g_{1}^{2}/{v_{g}}$ and $J_{2}=g_{2}^{2}/{v_{g}}$.
If there is no transmission and reflection observed by detectors
at the two ends of the waveguide, the state of system is projected
to $\xi_{k_{1}}|e_{1},g_{2}\rangle|0\rangle_{\rm{sp}}
+\xi_{k_{2}}|g_{1},e_{2}\rangle|0\rangle_{\rm{sp}}$,
which means that the entangled state between the two QDs has been generated.
The degree of entanglement can be
measured\cite{HorodeckiRMP2009, WoottersPRL1998} by the concurrence
$C = \rm{max}(\lambda_{1}-\lambda_{2}-\lambda_{3}-\lambda_{4},0)$.
$\lambda_{i}(i=1,2,3,4)$ are the square roots of the eigenvalues of the matrix
$R=\rho(\sigma_{y}\otimes\sigma_{y})\rho^{\ast}(\sigma_{y}\otimes\sigma_{y})$,
where $\rho$ is the density matrix of the system and $\sigma_{y}$ is the Pauli matrix.
For the system of two QDs, concurrence $C$ takes the form of
$C=\frac{2|\xi_{k_{1}}||\xi_{k_{2}}|}{|\xi_{k_{1}}|^{2}+|\xi_{k_{2}}|^{2}}$.
Maximum entanglement can be created when amplitude
$|\xi_{k_{1}}|$ is equal to $|\xi_{k_{2}}|$.

%- FIGURE 3 -%%
\setlength{\intextsep}{6pt plus 2pt minus 2pt}
\begin{figure}[htbp]
    \centering
    \includegraphics[width=6cm]{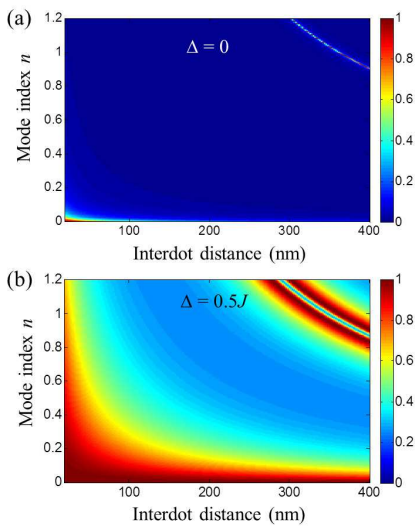}
    \vspace{-10pt}
    \caption{(Color online)
        Dependence of concurrence $C$ on the plasmonic
        waveguide mode index $n$ and the interdot distance $d$ for
        (a) on-resonance case $\Delta=0$ and
        (b) off-resonance case $\Delta=0.5J$.
        Here $\Gamma=0.01225J$ is used in the calculations.}
\end{figure}
%- END -%%

%%---%%
First, we examine the dependence of the concurrence on waveguide mode index,
interdot distance and QD-waveguide coupling strengths in the model.
Fig.~\blue{3} shows the concurrence $C$ as a function of the interdot
distance $d$ and the plasmonic waveguide mode index $n$ when the
coupling strength $g_{1}=g_{2}$.
We take the same detuning $\Delta=\delta$ between two QDs and the
waveguide mode (thus $J=J_{1}=J_{2}$) for simplicity in the following discussion.
Clearly, in Fig.~\blue{3(b)} when the incident waveguide mode is off-resonance
with QDs ($\Delta=0.5J$), the concurrence $C$ is higher than the
on-resonant case shown in Fig.~\blue{3(a)} over wide ranges of mode indices
and interdot distance.
For conventional waveguide modes with non-zero mode indices, maximum entanglement can be achieved
if the two QDs are placed at the right locations where the interdot
distance $d$ is equal to a multiple half-wavelength of the waveguide mode.
However, when the plasmonic waveguide mode index $n$ gets close to zero,
the concurrence maintains above $0.9$ over a wide range of interdot distance.
The relaxed distance requirement strongly overcomes the challenges of precise
QD position control in practical situations.

%- FIGURE 4 -%%
\setlength{\intextsep}{6pt plus 2pt minus 2pt}
\begin{figure}[htbp]
    \centering
    \includegraphics[width=6cm]{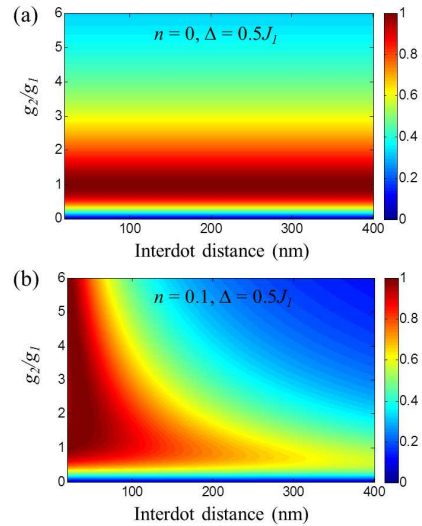}
    \vspace{-10pt}
    \caption{(Color online)
        Dependence of the concurrence $C$ on
        the $g_{2}/g_{1}$ ratio and the interdot distance $d$ when the
        plasmonic waveguide mode index (a) $n = 0$ and (b) $n = 0.1$.
        Here $\Gamma=0.01225J_{1}$ is used in the calculations.}
\end{figure}
%- END -%%

%%---%%
When the coupling strengths between QDs and the waveguide mode are not identical,
the dependence of the concurrence $C$ as a function of the ratio $g_{2}/g_{1}$ is
plotted in Fig.~\blue{4} when $\Delta = 0.5J_{1}$.
In Fig.~\blue{4(a)} when the mode index $n = 0$, maximum concurrence $C$ around unity can
be created for any arbitrary interdot distance when $g_{2} = g_{1}$ and the concurrence
decreases when the difference between $g_{2}$ and $g_{1}$ increases.
This phenomenon can be understood that the amplitudes $\xi_{k_{1}}$
and $\xi_{k_{2}}$ always have the same value in the case of $g_{1}=g_{2}$ and $n = 0$,
which results in the concurrence $C = 1$.
Even when the mode index cannot reach exact zero in practical situations,
for example $n = 0.1$, high concurrence $C$ can still be achieved within
relative broad ranges of interdot distance $d$ and $g_{2}/g_{1}$ ratio.
The dark red region in Fig.~\blue{4(b)} highlights that high concurrence can be
obtained as long as $d<100\,\mathrm{nm}$ and $1<g_{2}/g_{1}<4$,
which significantly relaxes the strict conditions required in conventional waveguides.
%%

%%---%%
Next, we discuss concurrence of the entanglement state between two QDs
in a practical $D = 110\,\mathrm{nm}$ $\mathrm{SiO}_{2}/\mathrm{Ag}$ waveguide
with dispersion illustrated in Fig.~\blue{2(b)}.
Fig.~\blue{5(a)} shows the concurrence $C$ when the waveguide mode index $n = 0.022$,
$0.164$, $0.462$, and $0.962$.
Near-zero mode index can be chosen by working near the cut-off wavelength of
the waveguide mode.
The corresponding group velocity $v_g$ are $0.82\times10^7\,\mathrm{m}/\mathrm{s}$,
$1.65\times10^7\,\mathrm{m}/\mathrm{s}$, $3.44\times10^7\,\mathrm{m}/\mathrm{s}$,
and $9.55\times10^7\,\mathrm{m}/\mathrm{s}$, respectively.
For other parameters of the QD-waveguide device, coupling strength $g_{1}=g_{2}=35\,\mathrm{GHz}$,
detuning $\Delta=0.5J$ with $J=J_{1}=J_{2}$, and total dissipation $\Gamma=500\,\mathrm{GHz}$.
When $n$ is $0.962$, the maximum value of concurrence occurs at two locations (green dash-dotted line in Fig.~\blue{5(a)}) where the interdot distance $d = 288\,\mathrm{nm}$ and $d = 312\,\mathrm{nm}$.
The special locations for creating high concurrence generally satisfy
$\Delta=-(\frac{4J+\Gamma}{2})\tan(kd)$ and $kd=\mathrm{m} \pi$ ($\mathrm{m}$ is an integer),
which results in $|\xi_{k_{1}}|$ = $|\xi_{k_{2}}|$.
However, when $n$ is $0.022$, high concurrence can be maintained for any interdot
distance $d$ over several hundreds of nanometers.
%5
This is due to the fact that phase variation along the waveguide is very small for the
plasmonic waveguide mode with near-zero index.
Moreover, in Fig.~\blue{5(b)}, the concurrence $C$ for various $\Delta$ from $0.1J$ to $0.5J$
is shown when the mode index $n$ is $0.022$.
Clearly, the high concurrence $C$ can be created for a wide range of the interdot
distance $d$ when $\Delta$ is $0.5J$.
This means that certain amount of QD-waveguide detuning in experiment will not be
detrimental to the high concurrence across large interdot distance
as long as the QD-waveguide coupling strength is maintained.
%%
%- FIGURE 5 -%%
\setlength{\intextsep}{6pt plus 2pt minus 2pt}
\begin{figure}
    \centering
    \includegraphics[width=5.5cm]{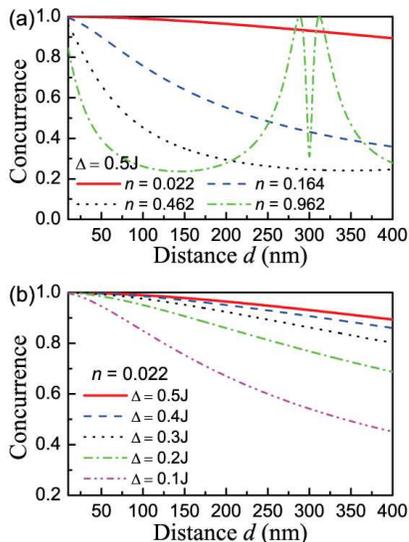}
    \vspace{-10pt}
    \caption{(Color online)
        (a) Dependence of the concurrence $C$ on
        the interdot distance $d$ when $\Delta = 0.5J$ for different mode
        index $n$ at $0.022$ (red solid line),
        $0.164$ (blue dashed line),
        $0.462$ (black dotted line), and
        $0.962$ (green dash-dotted line).
        (b) Dependence of the concurrence $C$ on the interdot distance $d$
        when the mode index $n = 0.022$ for different QD-waveguide detuning
        $\Delta$ at $0.5J$ (red solid line),
        $0.4J$ (blue dashed line),
        $0.3J$ (black dotted line),
        $0.2J$ (green dash-dotted line), and
        $0.1J$ (magenta dash-dot-dot line).}
\end{figure}
%- END -%%

%%---%%
In conclusion, we have examined quantum entanglement in $\mathrm{SiO}_{2}/\mathrm{Ag}$
plasmonic waveguides with considerations of QD-waveguide detunings, asymmetric coupling strengths and dissipations.
The waveguides can be designed to possess near-zero mode indices around the QDs transitions, and high concurrence
can be achieved between two QDs interacting with the plasmonic waveguide modes.
A wide range of interdot distance is allowed for achieving high concurrence due to
the near-zero phase variation along the waveguide, which shows advantages over the schemes implemented by dielectric
waveguides or metal nanowires where specific interdot distances are required.
The plasmonic waveguide with near-zero mode indices serve as a great platform for solid-state quantum optics and quantum
information processing.

%%---%%
This work was supported by Energy Research and Development Center at Missouri University of Science and Technology and the University of Missouri Research Board.

%%%%%%%%%%%%%%%%%%%%%%%%%%%%%%%%%%%
%%%%%%%%%%% References %%%%%%%%%%%%
%%%%%%%%%%%%%%%%%%%%%%%%%%%%%%%%%%%

\end{document}